# Wireless Communications and Mobile Computing

# Research Progress of News Recommendation Methods

Jing Qin

Northeastern University, Shenyang 110169, China.

Correspondence: annyproj@126.com

Abstract

Due to researchers' aim to study personalized recommendations for different business fields, the summary of recommendation methods in specific fields is of practical significance. News recommendation systems were the earliest research field regarding recommendation systems, and were also the earliest recommendation field to apply the collaborative filtering method. In addition, news is real-time and rich in content, which makes news recommendation methods more challenging than in other fields. Thus, this paper summarizes the research progress regarding news recommendation methods. From 2018 to 2020, developed news recommendation methods were mainly deep learning-based, attention-based, and knowledge graphs-based. As of 2020, there are many news recommendation methods that combine attention mechanisms and knowledge graphs. However, these methods were all developed based on basic methods (the collaborative filtering method, the content-based recommendation method, and a mixed recommendation method combining the two). In order to allow researchers to have a detailed understanding of the development process of news recommendation methods, the news recommendation methods surveyed in this paper, which cover nearly 10 years, are divided into three categories according to the abovementioned basic methods. Firstly, the paper introduces the basic ideas of each category of methods and then summarizes the recommendation methods that are combined with other methods based on each category of methods and according to the time sequence of research results. Finally, this paper also summarizes the challenges confronting news recommendation systems.

1. Introduction

A recommender system (RS) [1] is a tool that actively pushes services onto users, recommending user preference correlated information to users through the method of "mind-reading", and continuously updating recommended content according to changes to users' preference. A RS not only saves time for users to acquire information but also allows users to explore more potential interests. At the same time, RSs are also considered the most effective means of information exposure. News websites are important channels for people to capture information strictly related to their work and daily life. It is essential to provide the needed information promptly and accurately to the users. Taking the "Yahoo NEWS" website as an example, if a user reads a piece of news titled "South Americans marvel at total solar eclipse," the webpage will automatically show content related to the solar eclipse.

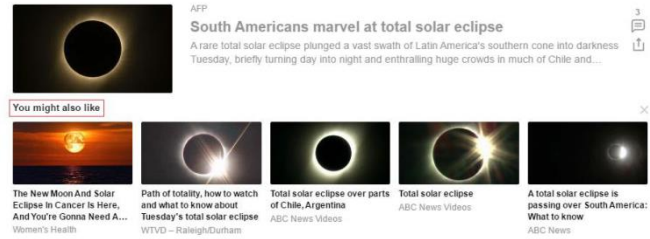

Figure 1: Recommendations after browsing a news item in Yahoo News.

The demonstration in Figure 1 only shows the recommended news content after non-registered users browsed a piece of news. If the users are registered to the website or are logged on to the news website on their mobile devices, they can get more news recommendations satisfying their personal preferences.

Due to different requirements in various business fields, researchers have proposed different recommendation methods for each type of field (e.g., e-commerce [2], travel [3], social networks [4], news [5]). Unlike other business fields, a news portal website has strong real-time performance, fast update speed, and a large quantity of content. Each news item contains a large amount of text content. News recommender systems first used the user-based collaborative filtering (UserCF) method [6], and then began using content-based (CB) recommendation methods [7], model-based collaborative filtering recommendation methods (e.g., matrix factorization method) [8], and hybrid CB and CF methods (e.g., recommendation methods based on deep learning [9], attention-based methods [10], knowledge graph (KG)-based methods [11]). In recent years, deep learning and knowledge graph methods have become hot technologies in the research field of news recommendation methods. In 2017, Park et al. [9] first used deep learning methods in the news recommendation field. In this method, it used recurrent neural network (RNN) and convolutional neural network (CNN) methods to obtain the user profile. In 2018, the authors of [10] first used the attention mechanism to realize user portrait in news recommendation. In the same year, the authors of [11] first used the knowledge graph to realize user profiles in the news recommendation field. Then in 2019 and 2020, researchers combined attention mechanism, knowledge graph, and deep learning methods to design news recommendation methods, thus improving the accuracy of recommendation results.

To provide researchers with an extensive overview of news recommender systems in recent years, this review outlines research published from 2011 to 2020, including some representative correlative papers about news recommendation methods. Unlike other reviews of news recommendation methods [5,12-15], this paper summarizes the research progress of news recommendation methods step by step, starting from the basic ideas of the most basic news recommendation methods and following the timeline of the development of recommendation methods.

The summary presented in this paper is structured into six steps: (1) the characteristics of news and the general news recommendation frameworks, (2) the basic idea of CF methods and their research progress regarding news recommendation methods, (3) the basic idea of content-based recommendation methods and the research progress in applying them to news recommendation methods, (4) the basic idea of combined CB and CF recommendation methods and the research progress in applying them to news recommendation methods, (5) evaluation metrics in news recommendation methods, and (6) the research prospects of news recommendation methods.

2. The Characteristics of News and the General News recommendation Framework

**2.1 The characteristics of news**

Compared with the items recommended by other recommender systems, such as those for e-commerce, travel, and videos, news is mainly characterized by strong timeliness, rich news content, and proximity.

(1) Strong Timeliness
News articles are a report of recent facts that are of social significance and arouse public interest. Timeliness is reflected in two aspects: recent occurrence and new content. Users are likely to pay more attention to current affairs news items.
(2) Rich News Content
News has a fixed structure, which generally includes five parts: title, lead, subject, background, and conclusion. To facilitate retrieval, editors of some news websites add keywords to news articles. In addition, each news content contains a large amount of topic information, such as sports, medical treatment, and music. The background in each news item also implies the information presented in the news. Therefore, the mining of news content is an important task of news recommendation methods.
(3) Proximity
News items are close to users geographically or in terms of thoughts and interests. The closer the location of the incident is to the user's location, and the more concerned the user is, the higher is the value of the news to the user. The closer the matter is to the vital interests of users, and the more attention users give to it, the higher is the value of news to users.

As news has the above characteristics, the challenges faced by news recommender systems include analyzing a large amount of news content in a short time, fully considering the timeliness and popularity of news when recommending news, and accurately mining news events. In addition, since most users of news websites are unregistered users, news recommender systems face a large number of users' cold-start problems.

**2.2 The general news recommendation framework**

In recent years, many recommendation methods have paid more attention to the analysis of news content due to the large amount of information contained in news content. When selecting news data analysis, not only is the news data itself on the website considered, but the use of auxiliary data (e.g., cross-domain data, knowledge base) to analyze news content is considered, as shown in Figure 2.

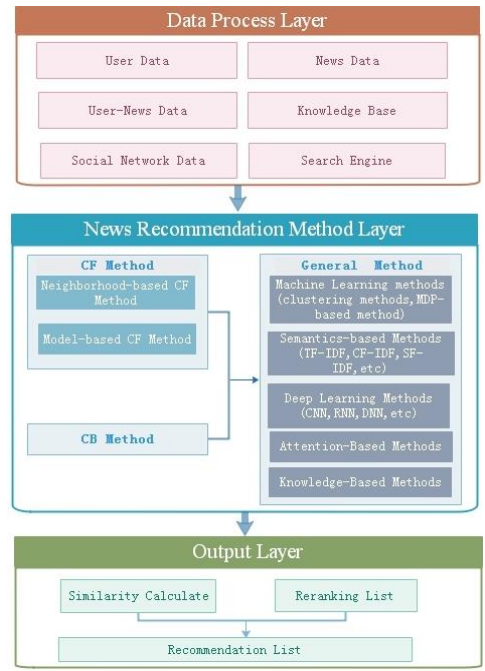

Figure 2: The general news recommendation framework.

The news recommendation framework is mainly divided into three parts: data processing layer, news recommendation model layer and output layer. The data processing layer contains the extracted and correlated data and word embedding process. The general data sources include user data (e.g., user registration information), news data (e.g., data on news websites), user browsing information data (e.g., user comments, browsing records), social network data (e.g., user friend data on social networks), knowledge base (e.g., Wikipedia, Freebase, Baidu Baike), and search engine tools (e.g., Bing search, Google search). The news recommendation model layer mainly uses semantic model, topic model, machine learning model, the deep learning model, and so on. In practical application, one or more models are combined to realize the task of news recommendation. The output layer generates the final recommendation list to the target user by calculating the similarity and re-ranking according to the prediction results of the model. The news vector representation and the user vector representation are also obtained from the learning-based model.

## 3. The Basic Idea and Progress of CF-Based News Recommendation Methods

This section will introduce the basic principle and recommendation process of the CF method, and then summarize the research progress regarding the CF method in the context of news recommendation.

### 3.1 The Basic Idea of CF Recommendation Methods

Goldberg [1] raised the concept of CF (i.e., people collaborating to help one another perform filtering by reading their reactions to documents they read) in 1992, and the idea of CF is applied in the mail filtering system (Tapestry). The authors of [6] proposed user-based collaborative filtering (UserCF) method, which was applied in Netnews, and this was the first news recommender system. The UserCF method is mainly used to predict users'rating of unread news in [6]. The UserCF method's rating prediction consists of three steps: (1) generating a user-item matrix using user rating data, (2) calculating the user similarity by Pearson correlation coefficients (Equation 1), and (3) predicting the user's rating for unread news by user similarity.

$$sim(u, v) = \frac{\sum_{i \in I}(R_{u,i} - \bar{R}_u)(R_{v,i} - \bar{R}_v)}{\sqrt{\sum_{i \in I}(R_{u,i} - \bar{R}_u)^2}\sqrt{\sum_{i \in I}(R_{v,i} - \bar{R}_v)^2}} \qquad (1)$$

where $R_{u,i}$ denotes the rating of user u for item i, $R_{v,i}$ denotes the rating of user v for item i, $\bar{R}_u$ is the average rating of the u-th users, and $\bar{R}_v$ is the average rating of the v-th users.

The authors of [16] proposed the item-based collaborative filtering (ItemCF) method. The ItemCF method is similar to UserCF, but ItemCF calculates the similarity between items. The authors of [16] used three similarity computation methods (cosine similarity (Equation 2), adjusted cosine similarity (Equation 3), and pearson correlation coefficients (Equation 4)).

$$sim(i, j) = cos(\vec{i}, \vec{j}) = \frac{\vec{i} \cdot \vec{j}}{\|\vec{i}\| * \|\vec{j}\|} \qquad (2)$$

where "." denotes the dot-product of the two vectors ($i$ and $j$). The adjusted cosine similarity formula is shown below.

$$sim(i, j) = \frac{\sum_{u \in U}(R_{u,i} - \bar{R}_u)(R_{u,j} - \bar{R}_u)}{\sqrt{\sum_{u \in U}(R_{u,i} - \bar{R}_u)^2}\sqrt{\sum_{i \in I}(R_{u,j} - \bar{R}_u)^2}} \qquad (3)$$

where $\bar{R}_u$ denotes the average of the u-th user's ratings. Pearson correlation similarity is given by

$$sim(i, j) = \frac{\sum_{u \in U}(R_{u,i} - \bar{R}_i)(R_{u,j} - \bar{R}_j)}{\sqrt{\sum_{u \in U}(R_{u,i} - \bar{R}_i)^2}\sqrt{\sum_{i \in I}(R_{u,j} - \bar{R}_j)^2}} \qquad (4)$$

where $R_{u,i}$ denotes the rating of user $u$ for item $i$, and $\bar{R}_i$ is the average rating of the i-th items. $R_{u,j}$ denotes the rating of user $u$ for item $j$, and $\bar{R}_j$ is the average rating of the j-th items.

Three kinds of similarity calculation method were compared, and the results show the mean absolute error (MAE) metric of adjusted cosine-based similarity is the lowest in [16]. However, it is still necessary to choose the similarity calculation method according to the specific dataset.

CF methods are divided into memory-based CF methods and model-based CF methods [17]. UserCF belongs to memory-based CF, and ItemCF belongs to model-based CF [14]. Besides ItemCF, model-based CF includes many methods, such as latent factor model (LFM) and machine learning-based methods. UserCF and ItemCF also are considered as neighborhood-based method [18]. The purpose of neighborhood-based methods is to calculate the relationship between the user or item [18]. The authors of [18] also consider the LFM as another important model-based CF method. The purpose of the LFM is to connect users' interests and items through latent factors.

The authors of [18] proposed a matrix factorization (MF) based recommendation method, and MF belongs to the model-based CF methods. The basic idea of the MF model is to decompose the user-item rating matrix into user latent factor vector and item latent factor vector in same latent factor space of dimensionality. Assume $\boldsymbol{R}^f$ denotes the latent factor space of dimensionality $\boldsymbol{f}$, and $\boldsymbol{q_i}$ denotes the factor vector of item $\boldsymbol{i}$, and $\boldsymbol{p_u}$ is the factor vector of user $\boldsymbol{u}$, and $\boldsymbol{q_i}, \boldsymbol{p_u} \in \boldsymbol{R}^f$. The basic MF model is shown in Equation (5)

$$\hat{\boldsymbol{r}}_{\boldsymbol{ui}} = \boldsymbol{q_i^T p_u} \tag{5}$$

where, $\hat{\boldsymbol{r}}_{\boldsymbol{ui}}$ denotes user u's rating of item $\boldsymbol{i}$. The challenge of the MF model is to obtain factor vectors ($\boldsymbol{q_i}, \boldsymbol{p_u}$) by a learning method [18] (e.g., (stochastic gradient descent, SGD), (alternating least squares, ALS)).

The main advantages and disadvantages of collaborative filtering (CF) methods are shown in Table 1.

Table 1. The main advantages and disadvantages of CF methods

| Method | Main Advantages | Main Disadvantages |
|--------|-----------------|--------------------|
| UserCF | 1. Fit for fewer users | 1. Does not consider the context of the item |
| | 2. Results are diverse | 2. Cold-start problem |
| | | 3. Results lack interpretability |
| ItemCF | 1. Fit for fewer items | 1. Does not consider the context of the item |
| | 2. Results are interpretable | 2. Results lack diversity |
| | | 3. Cold-start problem |
| MF | 1. Rating prediction is more accurate | 1 Does not consider the context of the item |
| | 2. Low space complexity | 2. Cold-start problem |
| | | 3. Results lack interpretability |

As can be seen from the table, each CF method has its advantages, but in the news field, due to the sparse data and fast news update speed, the direct use of the CF method for realizing news recommendation can no longer meet the needs of modern news recommender systems. Thus, in recent years, many improved methods were developed based on CF methods, especially using model-based CF methods.

## 3.2 Progress of CF-Based News Recommendation Methods

CF methods can recommend items for users only according to the user's scoring data. Due to the scarcity of users' scoring data in the context of news, only using CF methods will lead to low recommendation accuracy. Thus, in recent years, in order to improve the recommendation accuracy of CF methods, researchers have proposed a large number of hybrid methods based on CF methods. The particular methods include clustering-based methods [8, 19, 20, 21] and the Markov decision process (MDP)-based method [22].

### 3.2.1 Combining with clustering-based methods

The clustering method combined with CF includes news clustering and user clustering. The clustering method for news also takes into account the representation of news content. In order to introduce the integrity of the method, all news recommendation methods related to clustering methods are introduced in this section.

The authors of [8] used two kinds of CF methods (memory-based CF method and model-based CF method). The memory-based CF method uses items' co-visitation count. The model-based CF method uses clustering-based method ((probabilistic latent semantic indexing, PLSI) [23] and MinHash). The results of user clustering and the statistical results of news are stored in the user table (UT) and Story Table (ST) respectively. The UT and ST are stored in the Bigtable infrastructure [24]. The recommender system contains three components: the news statics server (NSS), the news personalization server (NPS) and the news front end (NFE). The NSS is used for updating statistics in the ST. The NPS is used for generating news articles recommendations. The NFE is responsible for obtaining user requests (recommend requests and update statics requests), and feeding back the results of user requests. When the NFE receives the recommend request, the process of recommendation is as shown in Figure 3.

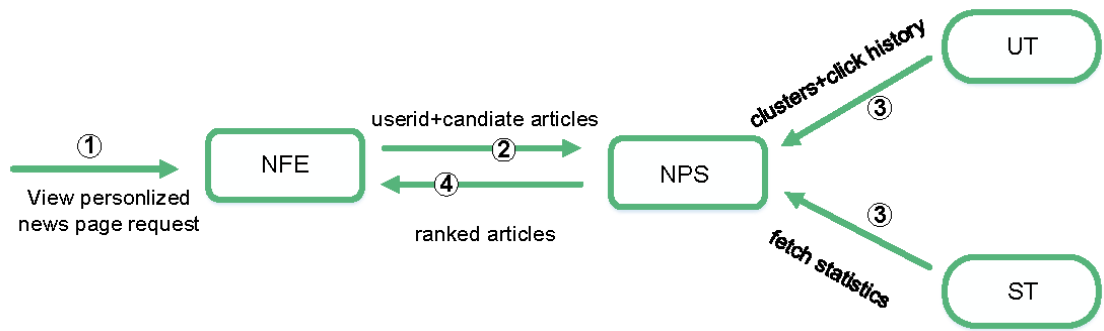

Figure 3: The process of recommendation (the NFE receives the recommend request) [8].

The authors of [19] proposed the scalable two-stage personalized news recommender system (SCENE). The recommendation framework of SCENE contains three components (news articles clustering, user profile construction, and personalized news recommendation). The news articles clustering component using locality sensitive hashing (LSH) [25] and hierarchical clustering to generate news groups. The user profile construction component combines accessed news content, similar access patterns and preferred name entities according to the user's history. The personalized news recommendation component compares the similarities between each small news group and the user's accessed news content, then recommends top-k news items to target users. The authors of [19] compared latent dirichlet allocation (LDA) [26] and PLSI for topic detection. LDA is suitable for a small news corpus, and PLSI is suitable for a larger news corpus.

The authors of [20] proposed a combine CF recommendation method with the user and news article cluster method based on WordNet-enabled k-means (W-kmeans) [27]. W-kmeans [27] uses the WordNet [28] database to cluster news articles in the user session. The authors of [20] proposed a recommendation architecture containing three components (session identification, clustering user sessions, and recommendation stage). The advantage of the proposed method is the use of word hypernyms with WordNet and clustering users by session instead of user's browsing history. The proposed method is better than the CF methods of latent semantic CF, and neighbor-based CF.

The authors of [21] proposed cluster-based CF recommendation methods. The recommendation methods focused on the news trends, and user preferences immediately changed. The recommendation architecture consists of three components (user modeling, clustering and evaluation, and personalized list generation). The user modeling component is used to calculate the user's vector representation using the word2vec [29] model and the user's vector representation is updated by users clicking on a news article. The clustering and evaluation model is used for user k-means clustering and calculate the click through rate (CTR). The personalized list generation model is used to generate a personalized news article list in real-time according to the user's request. In order to obtain fresh news articles, the paper proposed time decay function (TDF) and user time decay function (UTDF). The TDF calculates the scores of time decay according to the article's date of publication, and the UTDF calculates the scores of time decay according to the user's last access time. The study adopted mean average precision (MAP) and normalized discounted cumulative gain (NDCG) metrics [30] to evaluate the proposed method in offline experiments and online experiments. In the offline experiment, the performance of the proposed method was better than the compared clustering method (MinHash, Non-negative Matrix Factorization (NMF)) in the paper. In the online experiment, the performance of the UTDF was better than the TDF in CTR metric.

A summary of cluster-based CF methods is shown in Table 2.

Table 2. A summary of cluster-based CF news recommendation methods

| Ref. | Methods | Main Contribution |
|------|---------|-------------------|
| [8] | MinHash, PLSI | 1. fit for the massive scale of the datasets<br>2. support clustering over dynamic datasets |
| [19] | LSH, Hierarchical clustering | 1. use the name entities and news recency<br>2. support clustering on newly-published news articles |
| [20] | w-kmeans | 1. consider the word hypernyms with WordNet<br>2. clustering on session data |
| [21] | k-means | 1.fit for the massive scale of the datasets<br>2. fit for immediate news recommendation |

The combination of the CF method and the clustering method enhances the recommendation performance. However, there are some shortcomings to this combination, as follows: (1) the timeliness of news is not considered in clustering; (2) the context information of users is not considered; (3) It is hard to use clustering method for data with uneven categories; (4) the diversity of recommended news is only considered by the authors of [19], while the diversity is not considered in other methods.

### 3.2.2 Combining with the MDP-based method

The authors of [22] proposed an ItemCF with MDP method. The MDP method suitable for recommendation of news articles. Unlike ItemCF, the proposed method measures similarity of news items by semantic similarity [31]. The MDP uses four tuple representation: $< S, A, R_{wd}, TP >$, where $S$ denotes a set of states (i.e., each read news article as a state), A denotes a set of actions, $R_{wd}$ denotes the reward function, and $TP$ denotes the transition probability between states. This study defines three reward functions as follows:

$$Rwd_1(A_i, n_j, n_k) = \begin{cases} \frac{dt(n_k) - dt(n_j)}{N}, & trans_i(n_j, n_k) > 0 \\ 0, & otherwise \end{cases} \quad (6)$$

where, $trans_i(n_j, n_k)$ denotes the number of times $i$-1 articles were read from $n_j$ to $n_k$; $dt(n_k)$ and $dt(n_j)$ denotes the discovery time for news article $n_k$ and $n_j$ respectively.

$$Rwd_2(A_i, n_j, n_k) = \begin{cases} sim(n_j, n_k), & trans_i(n_j, n_k) > 0 \\ 0, & otherwise \end{cases} \quad (7)$$

where, $sim(n_j, n_k)$ denotes the semantic similarity measure [31].

$$Rwd_3(A_i, n_j, n_k) = \alpha * Rwd_1(A_i, n_j, n_k) + (1 - \alpha) * Rwd_2(A_i, n_j, n_k) \quad (8)$$

where, $\alpha$ is a hyperparameter ($\alpha$ is 0.7 is best in this experiment).

In this experiment, the proposed MDP methods are better than the other compared method (RegSVD [32], bigram [33], KNN [16]). Also, the performance of the $Rwd_3$ was better than the $Rwd_1$ and $Rwd_2$. However, when applying the MDP method, it is only related to the preamble state selected by the user, ignoring the complete history of the user, and at the same time, it does not consider the diversity of recommendation results.

4. The Basic Idea and Progress of News Recommendation Methods Based on CB

The content-based recommendation method was the important recommendation method in recommender system research. According to currently retrieved information, news recommendation system Newsweeder [7] was the first to use CB recommendation methods in 1995. CB recommendation methods are divided into text content statistics-based methods, semantics statistics-based methods, and methods use a mixture of both methods. There are fewer recommender systems that only use semantics statistics-based methods. Text content statistics-based news recommender systems include News Dude [34], Newsjunkie [35], YourNews [36]. News recommender systems based on the above mixed methods include OntoSeek [37], News@hand [38], etc.

**4.1 The Basic Idea of CB-based methods**

Text content statistics-based news recommendation methods often use vector space model (VSM) [39] to express keywords in every document. According to the news previously selected by users, this method can obtain users' unread news and similar news through calculating similarities between previously read news and unread news, and generate a recommendation list according to the result of similarity calculation. G.Salton [39] proposed the VSM expressing method in the 1960s. When news content uses VSM representation, every document is represented by n-dimension' space vectors, i.e., every dimension vector corresponds to a word in the given document's collection word list. Every document's content represents the vector, which is expressed by the keyword's weight. The weight refers to the relevance between the document and the keyword.

The conventional methods of calculating keyword weight include term frequency-inverse document frequency (TF-IDF) [40] and improved TF-IDF methods. The method of TF-IDF also considers inverse document frequency (IDF) [41] while it is calculating term frequency (TF). IDF [41] means the word frequently appears in one document but seldom appears in other documents.

In content-based news recommendation methods, the first step is to profile news data and user data; then to calculate the similarity between the profiles and the news previously selected by the user with the methods of cosine similarity, semantic similarity, hybrid similarity, etc.; and finally to recommend similar top-k news to users. When profiling news and user data, the news characteristics and user characteristics, which contain a large quantity of text, can be expressed as a character vector by the methods of TF-IDF, one-hot, word2vec, doc2vec [42], etc. The general recommendation process is as shown in Figure 4.

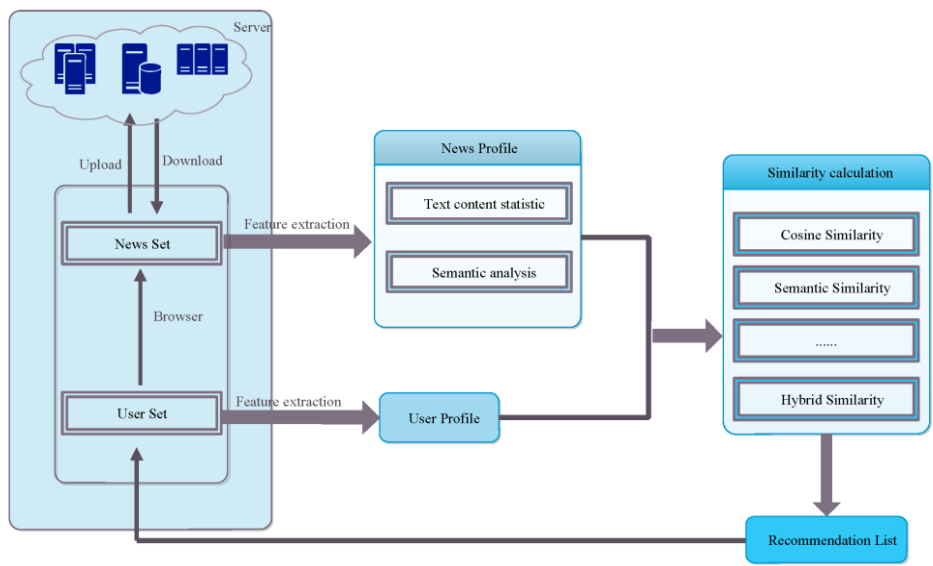

Figure 4: Recommendation process of CB news recommendation methods.

As is shown in Figure 4, CB news recommendation methods consist of two parts. One profiles the news and user, and the other calculates similarity after profiling. Researchers have also researched and improved news recommendation methods for these two aspects.

## 4.2 The Progress of CB News Recommendation Methods

Regarding content-based news recommendation methods, the research results in recent years have mainly focused on the VSM-based weight calculate methods and social network-based method.

### 4.2.1 VSM-Based Weight Calculate Methods

The core issue of studies on semantics-based news recommendation methods in recent years is to calculate weight value in VSM [39] expression. On the basis of the commonly used TF-IDF [40] method, many methods have been proposed, such as concept frequency-inverse document frequency (CF-IDF) [43], CF-IDF+ [44], synset frequence-inverse Document Frequency (SF-IDF) [45], Bing-SF-IDF [46], SF-IDF+ [47], Bing-SF-IDF+ [48], Bing-CF-IDF+ [49].

The CF-IDF method uses the concept in news content of replacing the characteristic word calculating weight, and the formula for calculating the CF-IDF weight is as follows.

$$cf - idf_{i,j} = cf_{i,j} \times idf_i \tag{9}$$

where $cf_{i,j}$ denotes the concept frequency, and $idf_i$ denotes the inverse document concept frequency, calculated by the following formulas:

$$cf_{i,j} = \frac{n_{i,j}}{\sum_k n_{k,j}} \tag{10}$$

where $n_{i,j}$ denotes the number of occurrences of a concept $C_i$ in document $D_j$, and $n_{k,j}$ denotes the total number of occurrences of all concepts in $D_j$.

$$idf_i = log \frac{|D|}{|\{d:c_i \in d\}|} \tag{11}$$

where $|D|$ denotes the total number of documents, and $d$ denotes the number of documents in which the concept $c_i$ appears.

The CF-IDF+ method adds the weight of the semantic relationship between concepts on the CF-IDF method basis. The formula for calculating the CF-IDF+ weight is as follows.

$$cf - idf_{i,j} = cf_{i,j} \times idf_i \times \omega_r \tag{12}$$

where $\omega_r$ denotes the weight of the semantic relationship. The CF-IDF+ method identifies three different weights (superclasses, subclasses, and domain relationships).

The Bing CF-IDF+ method adds page counts gathered by Web search engines for specific named entities. The Bing-CF-IDF+ method is a weighed combination of the Bing and the CF-IDF+ similarity measures, and the formula is as follows.

$$sim_{Bing-CF-IDF+}(u,d) = \alpha \times sim_{Bing}(u,d) + (1 - \alpha) \times sim_{CF-IDF+}(u,d) \tag{13}$$

where $u$ denotes the profile of the user u, and d denotes the unread news item $d$. $sim_{Bing}(u,d)$ denotes Bing similarity measures, and $sim_{CF-IDF+}(u,d)$ denotes $CF-IDF+$ similarity measures. $sim_{CF-IDF+}(u,d)$ uses the cosine similarity measure, normalized using a min-max scaling between 0 and 1. For Bing similarity measures, the formula is as follows.

$$sim_{Bing}(u,d) = \sum_{(u,d) \in V} log \left( \frac{P(u,d)}{N} / \left( \left( \frac{P(u)}{N} \right) \times \left( \frac{P(d)}{N} \right) \right) \right) / |V| \tag{14}$$

where $V$ denotes the set of named entity pairs (cartesian product of named entity set in user profile and named entity set in unread news items), $N$ denotes the total number of web pages, $p(u,d)$ denotes the Bing page count for pair$(u,d)$, $p(u)$ denotes the page counts for named entities $u$, and $p(d)$ denotes the page counts for named entities $d$.

The SF-IDF method is also similar to TF-IDF, as it uses the WordNet synonym sets. The SF-IDF formula is as follows.

$$sf - idf_{i,j} = sf_{i,j} \times idf_i \tag{15}$$

where $sf_{i,j}$ denotes the synset frequency, $idf_i$ denotes the inverse document synset frequency.

SF-IDF+ method adds the weight of semantic relationship between synset on SF-IDF method basis. The formula for calculating the SF-IDF+ weight is as follows.

$$sf - idf_{i,j} = sf_{i,j} \times idf_i \times \omega_r \tag{16}$$

where $\omega_r$ denotes the weight of the semantic relationship related synset.

The Bing SF-IDF method adds page counts gathered by Web search engines for specific named entities. The Bing-SF-IDF method is a weighed combination of the Bing and the SF-IDF similarity measures, and the formula is as follows.

$$sim_{Bing-SF-IDF}(u,d) = \alpha \times sim_{Bing}(u,d) + (1-\alpha) \times sim_{SF-IDF}(u,d) \tag{17}$$

where u denotes the profile of the user u, and d denotes the unread news item d. $sim_{Bing}(u,d)$ denotes Bing similarity measures, and $sim_{SF-IDF+}(u,d)$ denotes SF-IDF similarity measures. $sim_{SF-IDF+}(u,d)$ uses the cosine similarity measure, normalized using a min-max scaling between 0 and 1. The Bing-SF-IDF+ method is a weighed combination of the Bing and the SF-IDF+ similarity measures, and the formula as follows.

$$sim_{Bing-SF-IDF+}(u,d) = \alpha \times sim_{Bing}(u,d) + (1-\alpha) \times sim_{SF-IDF+}(u,d) \tag{18}$$

The development of these methods is shown in the Figure 5, in which, Figure. 5a depicts an improved process for the CF-IDF, CF-IDF+, and Bing-CF-IDF+ methods. Figure 5b depicts an improved process for the SF-IDF, SF-IDF+, and Bing-SF-IDF+ methods.
The authors of [48] made a comparison of some methods, such as Bing-SF-IDF+, SF-IDF, and TF-IDF. Bing-SF-IDF+ is better than the SF-IDF and TF-IDF methods in terms of F1-Score and Kappa statistics [50]. The authors of [49] made a comparison of some methods, such as Bing-CF-IDF+, TF-IDF, CF-IDF and CF-IDF+, which shows that Bing-CF-IDF+ is better than TF-IDF, CF-IDF and CF-IDF+ in terms of F1-score and Kappa statistic.

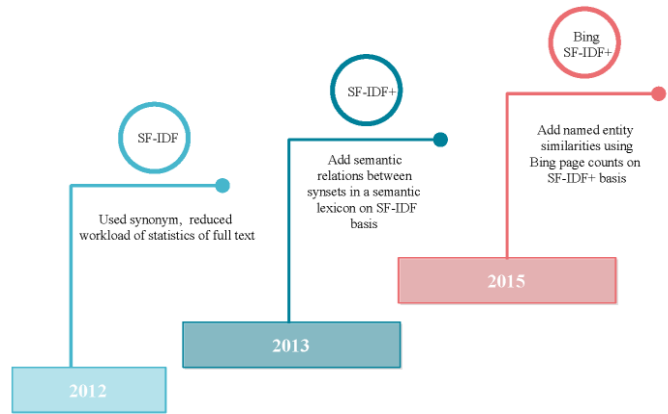

(a)

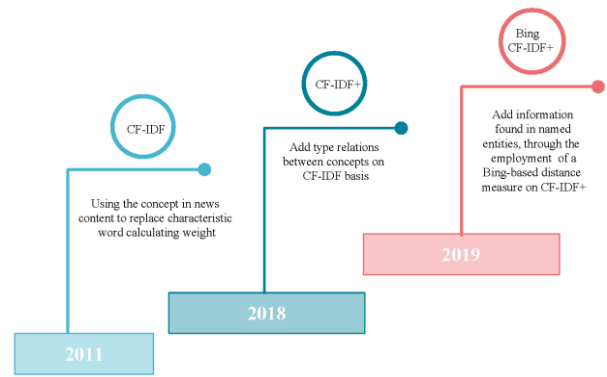

(b)

Figure 5: The improved process of the CF-IDF, CF-IDF+, Bing CF-IDF+ (a), The improved process of the SF-IDF, SF-IDF+, Bing-SF-IDF+ (b)

However, these methods also have their drawbacks, such as not deleting stop words, not considering the position information of particular words in the article, not considering synonyms, not considering the relationships between words, and a high dependency on lexicons. In addition, the diversity of recommended results has not been considered.

### 4.2.2 Combining with Social Networks-Based Methods

Hot news is usually related to hot topics on social networks. Twitter is a popular social platform where you can get user location information [51] and topic information from Twitter via its API (https://api.twitter.com/). Therefore, in the research of news recommendation methods, the information on Twitter is used to improve the accuracy of news recommendation. Use the user's Twitter information in [52, 53] to model the user profile and then recommend news articles for the user. There are three ways to model users using Twitter data (hashtag-based, topic-based, and entity-based). To verify the effectiveness of using Twitter data to recommend news articles, the authors constructed a data set by Twitter information and news articles. Then verify the influence of three different user modelling methods on the recommendation effect in the data set. Experiments show that the entity-based method has the best effect, but in the short-term recommendation, the topic-based method is better than the entity-based method.

The authors of [54] create a social news service. The social news service is a content-based method. Also, this method uses two content indexes (one from Twitter and one from RSS feeds). The authors of [55] use the popularity of topics on Twitter to sort news. The authors of [51] use the user's location information to find location-related topic information from Twitter. The authors of [56] propose fuzzy logic method for predicting users' interest. According to the user's interest category, virus news articles and their categories are analyzed by mining social media (Facebook and Twitter).

For the convenience of readers, the contents related to the methods of combining with social networks are summarized as follows (see Table 3).

Table 3. A summary of the social networks-based methods news recommendation methods.

| Ref. | Time | Main Methods | Social networks |
|------|------|--------------|-----------------|
| [52] | 2011 | Topic Model | Twitter |
| [54] | 2011 | Lucene indexes, TF-IDF | Twitter |
| [53] | 2013 | Topic Model | Twitter |
| [55] | 2013 | SOLR (in Apache Lucene project) | Twitter |
| [51] | 2018 | fuzzy clustering, naïve Bayes model | Twitter |
| [56] | 2020 | fuzzy logic method | Twitter, Facebook |

In addition, the authors of [57] proposed a neural social collaborative ranking (NSCR) approach, which seamlessly links users' information in the information domain with that in social networks. Regarding the information domain, it is used to strengthen users' and items' embedded learning. Regarding the social network domain, it is used to spread learning to non-bridge users through bridge users. Even though the NSCR method has not been tested on news datasets, it can still be applied to news recommendations.

In summary, the use of social networks can alleviate the problem of cold-start in news recommendations and recommend news according to popular trends in the news. However, the auxiliary information provided by social networks for news content analysis is limited, and cannot accurately analyze the entities in the news and the relationships between entities. In addition, in the above-mentioned news recommendation method based on social networks, the diversity of news recommendations and the domain knowledge related to the news are not considered.

5 The Progress of News Recommendation Methods Based on Combine CF and CB

CF method will bring about the problem of cold-start, especially when the data is sparse, the problem of cold-start is more serious. CB method can alleviate the problem of cold-start by analyzing news content. But in the CB method, the recommended results lack diversity (too similar to the history of users). In the news recommendation methods mixed with CB and CF methods, the CB method is used for the representation of news content, and CF methods are used to calculate the similarity of users or news or MF-based modelling. For example, the News Recommender System "Fab" introduced in 1997 [58].

In the methods mixing CF and CB, the main research results in recent years have focused on the methods of assisting recommendations in combination with search engine-based query methods, with the timeliness of news mining methods, with deep learning, attention mech, and KG. The CF method used in combination with deep learning, attention mechanism, and KG is a model-based method. The KG and attention-based methods were both applied to news recommendation research in 2018, and are also widely used at present.

## 5.1 Combining with search engine

The search engine is a way of enriching the news content and calculating the news similarity. The authors of [59, 60] use a search engine (enrich the news's content) combined with CF methods. The authors of [59] used methods of content-based collaborative filtering (CCF) and extended CCF (CCF+), which calculate the similarities in two document collections with the help of the Fisher Kernel function [61], and enrich search text' implication by using the Bing search engine. Through the experiment, the root mean square error (RMSE) metric performs better than Bayesian probabilistic matrix factorization (BPMF) method, and an effective content-based filtering method (Content). The authors of [60] proposed that using a search engine brings three advantages (short response time, fast processing of new content, and limited storage requirements).

Although search engines can quickly retrieve news content, there are still some drawbacks in the mining of news content, such as a lack of event analysis, entity analysis, and synonym analysis in news content. In addition, the reliability of the results found in search engines cannot be guaranteed.

## 5.2 Combining with the timeliness of news

Because of the timeliness of news content, the news life cycle is given attention in news recommendation methods.

The authors of [62] proposed a recommendation method of sports news. The RS has two components, one is CB-based component, and the other is CF-based component. In content-based component, only news articles in three days are recommended, and in the CF-based component, the older news articles are recommended. When analyzing the news article, 5 to 10 keywords can be extracted to represent the news article. The evaluation of results is good in KPI (Key Performance Indicators) and SUS (System Usability Scale).

The authors of [63] also proposed a hybrid CB and CF method, namely FeRe. In the CB part of the method, not only the news content but also the popularity and trends of the news are considered. For example, different types of news have different life cycles. Therefore, users have different preferences for different types of news. Some news users do not care when the news occurs but pay more attention to the content. In the CF part of the method, candidate news predictions are scored by calculating the similarity between users. Experiments showed that the recall rate of the FeRe method is better than those of either the CB method or CF method alone.

The authors of [64] proposed a session-based recommendation method, using the CF method and the CB method to calculate the similarity of news articles. The CF method uses the session-item matrix, in which the similarity of items is calculated as follows:

$$simCF(t_p, i, j) = \sum_{s \in s_{t_p,i} \cap s_{t_p,j}} (r_{s,i} \cdot r_{s,j}) / (\sqrt{\sum_{s \in s_{t_p,i}} (r_{s,i})^2} \sqrt{\sum_{s \in s_{t_p,j}} (r_{s,j})^2}) \qquad (19)$$

where $t_p$ denotes the time period $p$, each $t_p$ time period represents a session, $s_{t_p,i}$ denotes the session set of items $i$ in the specified time period $p$, and $s_{t_p,j}$ denotes the session set of item $j$ in the specified time period $p$. $r_{s,i}$ and $r_{s,j}$ respectively indicate whether items $i$ and $j$ are accessed in the session; if accessed, the result is 1, otherwise it is 0.

The content-based method uses the article category to calculate similarity, as follows:

$$simCB(t_p, i, j) = \sum_{c \in C_{t_p,i} \cap C_{t_p,j}} (f_{c,i} \cdot f_{c,j}) / (\sqrt{\sum_{c \in C_{t_p,i}} (f_{c,i})^2} \sqrt{\sum_{c \in C_{t_p,j}} (f_{c,j})^2}) \qquad (20)$$

where C denotes the set of topic categories, $f_{c,i}$ and $f_{c,j}$ respectively denote the probability that articles i and j belong to topic category c, and $C_{t_p,i}$ and $C_{t_p,j}$ respectively denote the topic category to which article i and j belong in time period t.

After combining Equations (19) and (20), the formula is as follows:

$$simCBCF(t_p, i, j) = \alpha * simCB(t_p, i, j) + (1 - \alpha) * simCF(t_p, i, j) \qquad (21)$$

where $\alpha$ represents the weight value, and 0, 0.5, and 1 were used in the experiment. The experimental results showed that the precision rate was the highest at $\alpha = 1$ and the lowest at $\alpha = 0$. However, when the timepoint value was small, i.e., $\alpha = 0.5$, it was similar to the precision rate achieved at $\alpha = 1$. Compared with the method based on the popularity baseline method, the average accuracy of the CBCF method was higher.

Some researchers also consider news popularity as a recommendation factor [65,66]. For example, the authors of [65] proposed an online recommendation method that combines trends and short-term user habits, takes the life of items into account, and divides news items into three classes (very popular, medium popular, unpopular). The authors of [66] proposed a based on most-popular strategies recommend methods online and offline in the CLEF NewsREEL challenge.

In summary, although the accuracy of the recommended results has improved in terms of news popularity and news life cycle, the following problems still exist: (1) the acquisition of users' complete preferences is ignored. For example, only the preferences in the current session are obtained [64], while in other methods, only the long-term preferences of users are obtained while ignoring the short-term preferences of users; (2) the connections between news content are ignored; and (3) the relationship between popular news and news that is important to users is ignored.

### 5.3 Combining with Deep Learning Methods

Based on the idea of MF, the authors of [67] firstly proposed neural collaborative filtering method. In 2017, deep learning methods were used in the news recommendation domain for the first time [9]. Compared with machine learning methods, deep learning methods' advantages are stronger learning ability and generalization ability. Since 2017, CF-based news recommendation methods used recurrent neural network (RNN) and convolutional neural network (CNN) models. The RNN model contains bidirectional RNN (Bi-RNN) and Long Short-Term Memory networks (LSTM), this model was used for modeling sequence data. The CNN model contains 1-D CNN, 2-D CNN, and 3-D CNN. The 1-D CNN similar to RNN, it mainly used for modeling sequence data. The 2-D CNN mainly used for modeling picture data, and the 3-D CNN mainly used for modeling video data. CNN also used for modeling the vector representation of news items in news recommendation methods. The authors of [9] used two different RNN models: one RNN model was used for modeling based on session data, and the other RNN model was used for modeling based on session and user click history. The study also used a CNN model to obtain users' long-term interest preferences.

The authors of [68] proposed a 3-D CNN-based news recommendation method. This model contains three parts (3-D tensor, 3-D convolutions, and score aggregation). The 3-D tensor part uses a three-dimensional structure to denote the similarity tensor. The proposed recommendation framework as follows (see Figure 6).

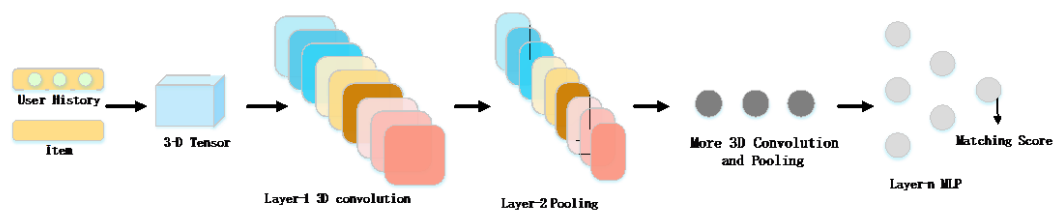

Figure 6: 3-D CNN-based news recommendation framework [68].

The authors of [69] also used the 3-D CNN model. The proposed model architecture includes User-History Component and Item Component components. The User-History Component used the 3-D CNN model, and the item Component used the 2-D CNN model. The authors of [70] not only uses user browsing information, but also uses the content information of news, and obtains article-embedding through the news information that users browsed. The recommendation model proposed in this paper includes two components (user profile component and article representation component). The user profile component is used to generate vector representation from news headlines and contents through doc2vec [42] embedding. The article representation component uses a method similar to MF, and uses a neural network to establish a model for calculating similarity between user preferences and candidate sets. The logistic function is directly used to generate the recommendation list of the target user during similarity calculation. The model architecture is shown in Figure 7.

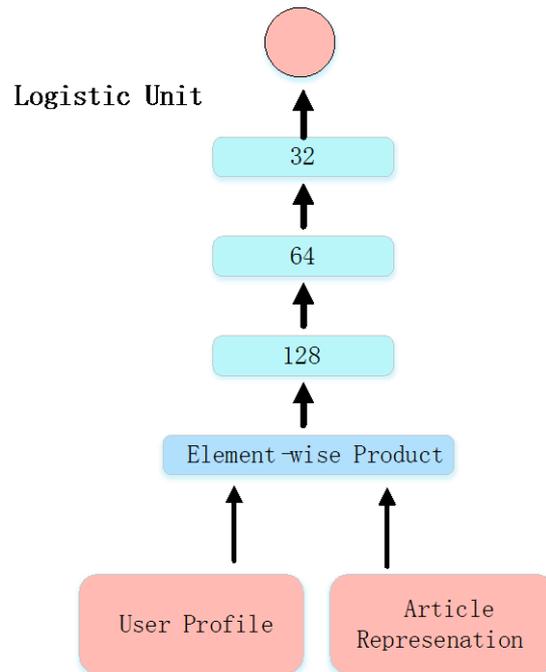

Figure 7: Neural Network Architecture [70].

The authors of [71] proposed the hybrid recurrent attention machine (HRAM) recommendation model, which also consists of two components. The first component uses the generalized matrix factorization method to predict the user will give the news articles. The second component is the user history component, which is implemented by an attention-based recurrent network. The model architecture is shown in Figure 8.

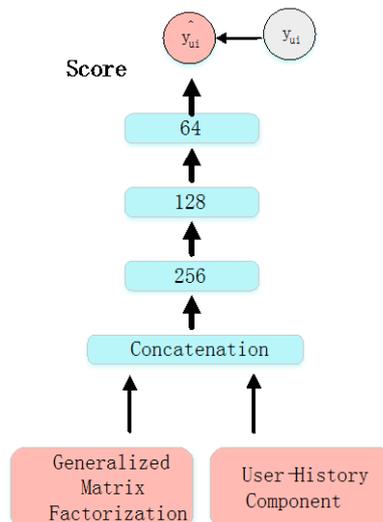

Figure 8: Hybrid Recommendation Attention Machine (HRAM) [71].

In the above-mentioned deep learning method, the traditional ID-based matrix decomposition method is replaced by the point multiplication method based on user characteristics and news characteristics, and a multi-layer neural network is added to make the model result more accurate than the traditional MF method. However, these methods still have some deficiencies: (1) they do not consider the timeliness of news (2) they lack diversity in recommendation results (3) they ignore the fine-grained representation of news content (4) they do not consider the upper and lower information of users.

## 5.4 Combining with Attention-Based Methods

Attention mechanisms are mainly used to mine the hot spots of users' interest preferences and obtain keywords from news articles. The recommendation methods based on attention mechanisms include one or more attention layers. In the news recommendation model, the news profile aspect uses a CNN, and the user profile aspect uses an RNN and then adds a different attention mechanism layer. The attention mechanism in the news profile aspect extracts the critical words, entities, events, etc. The attention mechanism in the user profile aspect extracts the hotspots of the user's internet.

The authors of [10] proposed the deep fusion model (DFM), which can be applied to item retrieval and item re-ranking. The DFM includes representation learning and an attention mechanism. In contrast to cold-start users who use knowledge from other domains, other users directly use the news domain. The model leverages an attention mechanism to alleviate the factors mentioned above and let the model dynamically combine knowledge from different domains. The proposed model performed better than compared models (logistic regression (LR), gradient boosting decision trees (GBDT), factorization machine (FM), SVDFeature, DeepFM, YoutubeRank [72], and DeepWide [73]) in terms of area under the curve (AUC) [74] and Logloss.

The authors of [75] proposed the deep attention neural network (DAN) model, which uses the two attention-based models. The attention-based RNN (ARNN) model is used to find the sequential features of users' clicks, and the attention-based neural network (ANN) model is used to discover features of users' interest. The DAN model includes three models (PCNN, ANN, and ARNN). The PCNN is used as a news representation extractor (combining CNN-based title embedding and the CNN-based entity of news embedding on the PCNN model), the ANN is used to extract user interest, and the ARNN is used as a sequential information extractor (with the LSTM model). Notably, the DAN model enriched the presentation of news articles compared to other methods (LibFM, DSSM [76], DeepWide, DeepFM, DMF, DKN [11]) on Adressa dataset [77] in the study.

The authors of [78] proposed a neural approach with personalized attention (NPA) model. The NPA model contains three modules (news encoder module, user encoder module, and click prediction module). The news encoder module aims to express the news. The news encoder module contains three sub-modules (word embedding, contextual representations, and mining the import words using attention mechanisms). The user encoder module is used to learn the representations of users from the representations of their clicked news. The click predictor module is used to predict target users' click scores on each candidate news item. The proposed method performed better than compared methods (LibFM; CNN; DSSM; WideDeep; DeepFM; DFM; DKN) in terms of AUC, mean reciprocal rank (MRR), and NDCG metrics.

The authors of [79] proposed a topic-aware news encoder, and a user encoder in the news encoder used a CNN and an attention network for representations of news titles. The news encoder contains three layers (word embedding expression layer, CNN network layer, and attention mechanism layer). The word embedding expression layer converts a news title from a word sequence into a vector sequence. The CNN network layer uses the CNN to learn contextual word representations by capturing local contexts. The attention mechanism layer selects important words in news titles to learn informative news representations. The user encoder uses the user's browser history and the attention network as a representation of the user. Also, this paper added a news topic classification model to predict the topic of news. The proposed method performed better than the compared methods (LibFM; CNN; DSSM; WideDeep; DeepFM; DFM (Deep Fusion Model); DKN; GRU) in terms of the AUC, MRR, and NDCG metrics.

The attention mechanism generates news vector representations more comprehensively than other content-based methods and can represent news in a fine-grained manner based on the word level and can better tap users' interests. However, the abovementioned news recommendation method based on attention mechanism still has the following problems: (1) ignoring the loss of information between each attention layer; (2) ignoring the differences between long-term and short-term preferences of users; (3) not considering the relationship between entities in news, which affects the accuracy of news recommendations; (4) not considering the popularity and timeliness of news; (5) not considering the context information of users, which affects the reading experience of users. Similarly, the diversity of recommendation results is not considered in the attention mechanism method.

## 5.5 Combining with Knowledge Graph-Based Methods

KG was proposed by Google in 2012. It was first applied to the field of news recommendation in 2018 and has also been a research hotspot in recent years [11]. KG is a method that represents knowledge by using a graph model. It is divided into general KG (FreeBase [80], Yago [81]), and domain KG (News Graph [82]). Many methods of using KG also combine attention mechanisms [82,83,84], thus being better suited for news vector representation.

The authors of [11] proposes the deep knowledge-aware network (DKN) framework that brings a knowledge graph into news recommendation and adds an attention module. The DKN framework is a content-based recommendation for CTR prediction, fit for highly time-sensitive news. This knowledge graph consists of the news content with a relevant entity. The DKN includes a critical component, namely knowledge-aware convolutional neural networks (KCNN), to fuse the word-level and knowledge-level representations of news and generate a knowledge-aware embedding vector. Experiment results indicate that the values of F-Measure and Area Under the Curve (AUC) were superior to those of other models compared (LibFM, a feature-based factorization model [85]; knowledge powered convolutional neural Network, KPCNN [86]; deep structured semantic model, DSSM; DeepWide; DeepFM [87]; YouTubeNet [72]; and deep matrix factorization model, DMF [88]).

The authors of [89] proposed a novel self-attention-based model. Compared with DKN model, the model uses word-level description instead of topic-level description. The attention model contains four-level self-attention modules (word-level, item-level, user-level, and multi-head attention). The word-level model transformed the words in news articles to word-based representations. The item-level model fused the representation among words, entities, and contexts. The user-level model was used to construct user-embedding. The multi-head attention model was used to concatenate history and candidate news, and then calculated the CTR via a fully connected layer. The method proposed is better than the compared methods (LR, DNN, DeepFM, DeepWide, DKN, Popularity) for the AUC metric and NDCG metric.

The authors of [82] proposed News Graph; it is the first time the knowledge graph was used in the news field to realize news recommendation tasks. News Graph is based on Microsoft Service Network (MSN) news data, and collaborative relations are added to the original KG, including Same News Triples, Same User Triples, and Same Session Triples. To verify the effectiveness of News Graph, four tasks were verified in the experiment (predicting the user's click on news items, news classification, news popularity, and the matching degree between the locations of news items and the interest of local users). The experimental results showed that News Graph performed better than the original KG in the above tasks in accuracy, AUC, and F1-score metrics.

The authors of [90] proposed the SED (short distance over knowledge graphs) method, which consists of three steps: (1) finding the named entities involved in the news in FreeBase, and then finding the nodes connected by these entities; (2) generating subgraphs according to the nodes found in the first step; and (3) finding similar news in the subgraph by calculating the shortest distance between entities. In addition, CNRec, a public dataset for KG research was proposed by the authors of [90] (https://github.com/kevinj22/CNRec). The SED method can alleviate the cold-start problem, and the combination of the SED method with the traditional TF-IDF and doc2vec methods can also improve the recommendation performance.

The authors of [83] proposed the topic-enriched knowledge graph recommender system (TEKGR), which improves upon the existing KG methods by mining external knowledge in news entities and related topics. The TEKGR framework is divided into three layers (using KG to model news headlines with three coding methods, obtaining users' interest expressions through attention mechanisms, and using the dot production method to calculate users' scores on candidate news). Experiments showed that TEKGR is superior to comparative methods (DKN, DeepFM, LibFM, LSTUR [91]) in AUC and F1-score metrics.

The authors of [84] proposed a graph enhanced representation learning (GERL) approach, which is a mixture of graph representation and attention mechanisms. GERL not only models users and news but also considers the neighbors of users and news, thus increasing the learning ability between users and news. In addition, the transformer [92] model is applied when generating news vector representations. Experiments showed that the GERL method is superior to DKN, LibFM, DSSM, and other methods according to AUC, MRR, and NDCG metrics.

The authors of [93] proposed a node2vec-based [94] economic news recommendation model. This method is mainly used to recommend economic news related to stocks for users, and the timeliness of news is considered in the model. In this paper, a securities KG was established, which includes stock, industry knowledge, related concepts, and other contents. Experiments showed that the model is superior to the traditional CF and CB recommendation methods, and can recommend personalized economic news for users in real-time in practical applications.

The KG can compensate for the cold-start problem in the recommender system, as well as digging up more potential news information. Therefore, the accuracy of recommendation results is improved, but the abovementioned news recommendation method based on KG still has the following shortcoming: (1) the continuity of news events is not considered; (2) there are too many named entities and link nodes in the KG, and important entities and links are not considered, which makes the method more suitable for recommendation applications in the online environment; (3) the creation of KG does not consider cross-language domain knowledge, and KG cannot be directly used to complete cross-language recommendations; (4) in the field of news, there are not many domain KGs. For example, a KG specifically used in news recommendation was created by the authors of [82].

## 6 Evaluation Metrics in News Recommendation Methods

The evaluation of methods by research papers in recent years mainly concentrated on the precision of recommendation methods, the effectiveness of recommendation list ranking, the effectiveness of real-time recommendation and the application effect of recommendation. In these metrics, except for root mean squared error (RMSE) and mean absolute deviation (MAE), the larger the value of the other metrics, the better the recommendation effect.

The evaluation metrics of the recommendation's precision [95] include precision, recall, hit rate (HR), F-Measure (or F-Score), accuracy, RMSE, and MAE. Precision indicates the probability of users' interest in the news listed in the recommendation list. Recall indicates the probability of news that users are interested in to be recommended. HR indicates the total number of correct news recommendation items in each user's top-k recommendation list. It is an evaluation metric of recall rate used in top-k recommendation methods. F-Measure (or F-Score) metric comprehensively considers precision and recall. The accuracy metric indicates the ratio between numbers of correct recommended news items and that of all news samples. RMSE and MAE are used to forecast the score that users give for news, for example, the MF-based recommendation method.

The evaluation metrics of recommendation list ranking effectiveness includes normalized discounted cumulative gain (NDCG), mean average precision (MAP), and area under the curve (AUC) [74,96]. NDCG is used to indicate the relevance between users and recommended news items. MAP is similar to NDCG, but in the MAP metric, the relevance between users and recommended news items is a binary correlation. AUC measures the probability that a recommender system ranks a positive user-item interaction higher than negative ones.

The evaluation metrics of the practical application of real-time recommendation includes KPIs (Key Performance Indicators) and SUS (System Usability Scale) in [62]. KPIs metric used to evaluate if the recommender system is in good shape or not. SUS metric used to evaluate the usability of the recommender system.

The evaluation metrics of the application effect of recommendation include the click through rate (CTR) metric, Kappa statistic metric [50], and mean reciprocal rank (MRR) metric [74]. The CTR metric is used to evaluate the number of hits on the news recommended. The Kappa statistic is a metric to measure whether the proposed classifications made are better than those made by a random guess. The MRR metric is used to evaluate the advantages and disadvantages of a recommendation list.

## 7 Prospects for News Recommendation Methods

It can be seen that in the research of the above-mentioned news recommendation methods, researchers mainly focus on personalized recommendation and recommendation accuracy. However, there are few studies on user fatigue, non-personalized news recommendation, privacy protection, news diversity, and other aspects, which are still important issues in future news recommendation research.

## 7.1 User Fatigue in News Recommendation

The news recommended for users in the news recommender system is news similar to that in the user's browsing history, which makes users tired when reading news, thus affecting the user's reading experience. However, the factor of user fatigue is rarely considered in news recommendation methods. User fatigue is defined as the loss of interest of users in repeatedly recommended items, and CTR of news items is used to measure user fatigue [97]. However, the authors of [97] only consider the user's gender, age, and CTR of news items, and does not consider the semantic relationship of news items, users' long-term and short-term interests, etc. How to measure the factors of users' reading fatigue is also a problem worth studying. In addition, the diversity and serendipity aspects of relieving users' reading fatigue can be further studied in news recommendation methods.

## 7.2 Privacy Protection in News Recommendations

Most news recommendation methods are based on the historical behavior of users. Thus, these methods ignore the privacy protection of users' behaviors. The authors of [98] pointed out that there are relatively few recommendation methods based on privacy protection in the field of news recommendation, but it is also a problem worth studying in the field of news. In 2020, the authors of [99] proposed a news recommendation method based on local differential privacy (LDP) [100]. The purpose of this method is to save the user's behavior only in the user's device, instead of uploading the user data to the server. The portrait of the user is not a portrait of a user but a group portrait of the user, thus making up for the impact of the lack of user information on the accuracy of the portrait of the user.

## 7.3 Non-personalized in News Recommendations

The authors of [101] proposed a non-personalized news recommendation method based on future-impact. The purpose of this method is to provide users with news items that will become hot spots (news items that users will pay more attention to) in the future. The non-personalized news recommendation method can not only alleviate the cold start problem, but also reduce the complexity of the recommendation method in online news recommendation. However, non-personalized recommendation does not consider users' preferences and also affects the experience of some users. Thus, the combination of non-personalized recommendation and personalized recommendation methods is also a research direction for future news recommendation methods.

## 7.4 Recommendation of Long-Tail News

Long-tail news refers to news that is rarely recommended. Due to users rarely browsing long-tail news, long-tail news browsing records are few. In the recommendation model, long-tail news rarely becomes recommended news for users. However, long-tail news is also valuable and needs to be mined by the recommender system and recommend to the target users. Due to the timeliness of news, it is hard to mine long-tail news that meets users' preferences in a timely and accurate manner. Therefore, in future news recommendation methods, the recommendation model of long-tail news is an important research direction, which can be modeled using KG or attention mechanisms.

## 7.5 Recommendation of High-Quality News

In the news recommendation method, recommend results only consider the accuracy rate and do not consider the news' quality. High-quality news refers to the authenticity, refinement, and integrity of the content. There are usually many news reports on the same event, and it is difficult to find the best news to recommend to users. Taking current affairs news as an example, high-quality news is refined and authoritative. Mining high-quality news content is closely related to the choice of news content analysis methods. In the future, the recommendation of high-quality news will consider narrowing the scope of candidate news (i.e., analyzing news content after classifying news according to events and removing fake news), analyzing news content on social media, and analyzing news content with expert knowledge.

## 8. Conclusions

The emergence of the news recommender system reduces the time for users to find news, and can timely and accurately recommend the news of interest to users. The research of news recommendation system is a hot spot in the research field of recommender system. At first, we introduced the characteristics of news and the general news recommendation framework, then analyzed the research progress of news recommendation methods, finally summarized evaluation metrics in news recommendation methods, and discussed the prospects for news recommendation methods. We hoped to be of all help to researchers in related fields.

## Conflicts of Interest

The author declares that there is no conflict of interest regarding the publication of this paper.

## Funding Statement

This research received no external funding.

## Acknowledgments

The author would like to thank the authors of all the references.